\documentclass[AMA,STIX1COL]{WileyNJD-v2-modified}

\DeclareMathOperator{\diag}{diag}

\usepackage{setspace}
\usepackage{caption}
\usepackage{subcaption}
\usepackage{hyperref}  
\usepackage{scalerel}

\articletype{R E S E A R C H \ A R T I C L E}

\received{26 April 2016}
\revised{6 June 2016}
\accepted{6 June 2016}

\begin{document}

\title{Multiple multi-sample testing under arbitrary covariance dependency}

\author[1]{Vladimir Vutov}

\author[1]{Thorsten Dickhaus*}


\authormark{V.Vutov and T.Dickhaus: Multiple multi-sample testing under arbitrary covariance dependency}

\address[1]{\orgdiv{Institute for Statistics}, \orgname{University of Bremen}, \orgaddress{\state{Bremen}, \country{Germany}}}



\corres{*Thorsten Dickhaus, Institute for Statistics,  University of Bremen, \email{dickhaus@uni-bremen.de}}

\presentaddress{Institute for Statistics, University of Bremen, Bibliothekstr. 1, 28359 Bremen, Germany
}

\abstract[Abstract]{
Modern high-throughput biomedical devices routinely produce data on a large scale, and the analysis of high-dimensional datasets has become commonplace in biomedical studies. However, given thousands or tens of thousands of measured variables in these datasets, extracting meaningful features poses a challenge.
In this article, we propose a procedure to evaluate the strength of the associations between a nominal (categorical) response variable and multiple features simultaneously. Specifically, we propose a framework of large-scale multiple testing under arbitrary correlation dependency among test statistics. First, marginal multinomial regressions are performed for each feature individually. Second, we use an approach of multiple marginal models for each baseline-category pair to establish asymptotic joint normality of the stacked vector of the marginal multinomial regression coefficients. Third, we estimate the (limiting) covariance matrix between the estimated coefficients from all marginal models. Finally, our approach approximates the realized false discovery proportion of a thresholding procedure for the marginal p-values, for each baseline-category pair.
The proposed approach offers a sensible trade-off between the expected numbers of true and false rejections.
Furthermore, we demonstrate a practical application of the method on hyperspectral imaging data. This dataset is obtained by a matrix-assisted laser desorption/ionization
(MALDI) instrument. MALDI demonstrates tremendous potential
for clinical diagnosis, particularly for cancer research. In our application, the nominal response categories represent cancer subtypes. }


\jnlcitation{\cname{%
\author{Vutov and  Dickhaus}
\author{}} (\cyear{2022}), 
\ctitle{Multiple k-sample testing under arbitrary covariance dependency.} \cjournal{.}, \cvol{2017;00:1--6}.}
\maketitle


\section{Introduction}\label{sec1}
Many datasets in various scientific disciplines, such as neuroimaging, genomics, brain-computer interfacing, and others, are nowadays high-dimensional. Such datasets are frequently analysed by means of large-scale multiple testing for some phenotype \textendash \, for example, cancer subtypes \textendash \, on each feature among thousands of features. This approach has multiple applications in the life sciences.\cite{dickhausbook} At least since the seminal publication \cite{BH}, control of the false discovery rate (FDR) is a commonly used type I error criterion in high-dimensional multiple test problems. However, applying the well-known Benjamini-Hochberg \cite{BH} (henceforth, B-H) procedure or Storey's \cite{storey2002,storey2004} procedure, which have originally been designed for stochastically independent or weakly dependent p-values, respectively, can lead to an FDR inflation under certain forms of (strong) dependencies. \citep*{efron2007, efron2010, pfa, leek}
Under the assumption of positive regression dependency on subsets (PRDS), the B-H procedure still controls the FDR\cite{BY}, but checking the validity of the PRDS assumption on the basis of a sample is far from trivial and only established for a limited number of special cases like, for example, elliptically distributed vectors of test statistics.\cite{Bodnar-Dickhaus2017}  

For these reasons and with the goal of optimizing the statistical power of the multiple test under the constraint of (at least asymptotic) FDR control at the desired level, multivariate (FDR-controlling) multiple tests have drawn a lot of attention over recent years.\cite{handbook-chapter} A multivariate multiple test incorporates the dependence structure among test statistics or p-values, respectively, or an estimate thereof, explicitly in its decision rule.  
Large-scale multiple testing under general and strong dependency remains challenging and an active research topic in modern statistics. \cite{pfa, leek, sun, friquet2009} Among other approaches like, for instance, considering block dependencies in genetics\cite{madam,MVCHS}, several multi-factor models have been proposed to model dependencies among test statistics.\cite{pfa, leek, friquet2009, efron2007, efron2010} In particular, a general framework to approximate the false discovery proportion (FDP) has been introduced.\cite{pfa,pfa_poet} It employs a so-called principal factor approximation (PFA) of the (estimated limiting) covariance matrix of test statistics, assuming that these test statistics are (at least asymptotically for large sample sizes) jointly normally distributed. The general strategy of the approach is to apply a spectral decomposition of the aforementioned covariance matrix and then to deduct its principal factors that mostly induce the strong correlation dependency. Since the FDR is the expected value of the FDP, bounding the FDP (with high probability) ensures FDR control, too.

Biological datasets generated by advanced high-throughput devices typically contain thousands of measured variables, many of which are related.   
In this work, we consider a dataset obtained by a matrix-assisted laser desorption/ionization (MALDI) imaging mass spectrometry (IMS) tool, also known as MALDI imaging. MALDI has advanced considerably and demonstrates immense potential in numerous pathological applications. \cite{krieg, boskamp} MALDI-generated datasets contain thousands of measured variables (corresponding to molecular masses), many of which are (highly) related. \cite{johannes} 
 Since MALDI data do in general not exhibit dependencies in blocks, it has recently been advocated to model dependencies among MALDI features by means of factor models, and to employ PFA in the context of multiple two-sample comparisons based on MALDI data.\cite{vutovdickhaus}

In this work, we deal with multiple multi-sample hypotheses testing under arbitrary correlation dependency, in particular in the case of more than two samples or categories, respectively. We do so under the scope of multiple marginal multinomial regression models and by applying the PFA methodology. We call our method multinomial - principal factor approximation ("Multi-PFA"). 

The remainder of this article is structured as follows. In Section \ref{sec2}, we present the proposed multi-sample multiple testing approach. Section \ref{sec4} is dedicated to a simulation study under different data-generating scenarios. We demonstrate the practical application of the proposed approach to hyperspectral data in Section \ref{sec-MALDI-data}, and we conclude with a discussion and an outlook in Section \ref{sec-discussion}.

\section{Proposed Methodology}\label{sec2}
\subsection{Data Structures}
MALDI imaging data are typically stored in an $n \times p$ matrix $X = (x_{ij})_{\substack{1 \leq i \leq n \\ 1 \leq j \leq p}}$, where rows correspond to  mass spectra and columns correspond to thousands of mass-to-charge (m/z) values.\cite{alex2} Since MALDI data represent molecular masses of ionizable molecules, $X \in \mathbb{R}^{n \times p}_{\geq 0}$. However, our proposed framework makes no explicit assumption that data points have to be non-negative.
We approach the biological challenge of analyzing the associations between (individual) m/z values (predictors $(X_j)_{1 \leq j \leq p}$) and several cancer subtypes (categorical response $Y$) by carrying out multiple statistical hypothesis tests simultaneously. Under the $j$-th null hypothesis $H_{0j}$, there is no association between $X_{j}$ and $Y$, and we aim at testing the $p$ null hypotheses $H_{01}, \ldots, H_{0p}$ (against their two-sided alternatives $H_{11}, \ldots, H_{1p}$) simultaneously based on one and the same dataset. Throughout the remainder, we assume that $Y$ is a (random) nominal outcome variable, meaning that the response categories are unordered. Hence, for each $j \in \{1, \ldots, p\}$, the tuple $(X_{j}, Y)$ takes its values in  $\mathbb{R}_{\geq 0} \times \{1, 2, \ldots, q\}$, where we denote by $q$ the number of (response) categories.

\subsection{Marginal Modelling for Categorical Responses}
First, we model marginal associations for each predictor $X_j$ individually. The reason for considering marginal modelling relies on the fact that we consider an inferential framework, in which the number of covariates is either quite large or even (much) larger than the number of observational units ($n << p$). Thus, a model that includes all predictors in a single model cannot be fitted (reliably). 
In the context of large-scale multiple (association) testing, marginal linear regression models have been adopted to test an individual hypothesis for each predictor. \cite{ecdf, pfa} Marginal generalized linear models (GLMs), for positive outcome variables, have been considered in situations where the focus of the statistical inference is on the mean of the response given a group of predictors. \cite{basu} Also, marginal logistic regressions have been utilized to analyse child obesity \cite{pepe}, with the usage of generalized estimating equations for jointly estimating (univariate) associations. \cite{gee}

For this section, let $j \in \{1, \ldots, p\}$ be arbitrary, but fixed.  To model the conditional distribution $\mathbb{P}(Y| X_j)$, we assume a marginal multinomial regression with a canonical (logit) link function.
In general, the $q$ response categories cannot be treated as a one-dimensional response, and we have to introduce a dummy variable for each but one category. Therefore, we have a multivariate outcome variable in the (relevant) case of $q > 2$ (for more details, see Chapter 7 in \cite{agresti}). One (response) category is used as a baseline, and the multinomial logistic model is established by pairing each of the other nominal response categories with the baseline. Throughout this article, response category $q$ is used as the baseline. Furthermore, we mainly consider the case of $q=3$ throughout the remainder, for concreteness and because of its relevance for the application that we are going to present in Section \ref{sec-MALDI-data}. For each (marginal) fit, we consider intercept terms $\alpha_{jc}$ and slope parameters $\beta_{jc}$, for $ 1 \leq c \leq q-1$. 
We denote the observables (regressor and response variable) for the $j$-th marginal regression fit by $(X_{j}^{(i)}, Y^{(i)})_{1 \leq i \leq n}$, and we assume that these tuples are stochastically independent and identically distributed bivariate random vectors, each having the same (joint) distribution as $(X_{j}, Y)$.
For observational unit $i \in \{1, \ldots, n\}$, our model equation for the $c$-th baseline-category logit is given by
\begin{equation}
\label{logit}
   \log\left(\mathbb{P}(Y_{c}^{(i)} = 1 | X_j^{(i)}) / \mathbb{P}(Y_{q}^{(i)} = 1 | X_j^{(i)})\right) =: \log\left(\pi_{jc}^{(i)} / \pi_{jq}^{(i)}\right) = \alpha_{jc} + \beta_{jc} X_{j}^{(i)},
\end{equation}
where $Y_{c}^{(i)}$ denotes the dummy indicator for category $c \in \{1, \ldots, q-1\}$ pertaining to observational unit $i \in \{1, \ldots, n\}$.
 The unknown model parameters $(\alpha_{jc}$, $\beta_{jc})_{1 \leq c \leq q-1}$  can be estimated by the maximum (log-) likelihood (ML) principle. To this end, we follow the derivations in Section 7.1.4 of \cite{agresti}. Given that $\pi_{jq}^{(i)} = 1 - \pi_{j1}^{(i)} - \pi_{j2}^{(i)} - \ldots - \pi_{j,q-1}^{(i)}$ as well as $Y_q^{(i)} = 1 - Y_1^{(i)} - Y_2^{(i)} - \ldots - Y_{q-1}^{(i)}$, the log-likelihood contribution for the $i$-th observational unit is under the model specified in \eqref{logit} given by
\begin{align}
\label{ml}
    \log \left( \prod_{c = 1}^{q} \left\{\pi_{jc}^{(i)}\right\}^{Y_{c}^{(i)}} \right) &= \sum_{c = 1}^{q-1} Y_{c}^{(i)} \log \pi_{jc}^{(i)} + \left( 1 - \sum_{c = 1}^{q-1} Y_{c}^{(i)} \right) \log \left[ 1 - \sum_{c = 1}^{q-1}  \pi_{jc}^{(i)} \right] \nonumber  \\ 
    &= \sum_{c = 1}^{q-1} Y_{c}^{(i)} \log \frac{\pi_{jc}^{(i)}}{1 - \sum_{c = 1}^{q-1}  \pi_{jc}^{(i)}} + \log \left[1 - \sum_{c = 1}^{q-1} \pi_{jc}^{(i)} \right].
\end{align}
\\
Substituting  $\log(\pi_{jc}^{(i)} / \pi_{jq}^{(i)}) = \alpha_{jc} + X^{(i)}_{j}\beta_{jc}$ as well as $\pi_{jq}^{(i)} = 1/ \{1 + \sum_{c = 1}^{q-1} \exp(\alpha_{jc} + X^{(i)}_{j}\beta_{jc}) \}$ in \eqref{ml}, we find that
\begin{equation}
\label{mle}
 \ell\left(\hat{\alpha}_{jc}, \hat{\beta}_{jc} | Y^{(i)}, X_{j}^{(i)}\right) = \max_{\alpha_{jc},\beta_{jc}} \sum_{i = 1}^{n} \bigg\{ \sum_{c = 1}^{q-1} Y_{c}^{(i)}(\alpha_{jc} + X_{j}^{(i)}\beta_{jc}) - \log \Big[ 1 + \sum_{c = 1}^{q-1} \exp(\alpha_{jc} + X^{(i)}_{j} \beta_{jc}) \Big] \bigg\},
\end{equation}
where $\ell$ stands for the log-likelihood function, and $\hat{\alpha}_{jc}$ and $\hat{\beta}_{jc}$ denote the ML estimators. In practice, the estimation is carried out conditionally to the actual observed values of the response indicators and the regressors, leading to the numerical values of the ML estimates. 

In this work, we are interested in testing simultaneously two families of hypothesis-alternative pairs, which are given by
\begin{equation}
\label{testing}
H_{0jc}: \beta_{jc} = 0 \text{~~versus~~} H_{1jc}: \beta_{jc} \neq 0, \quad j = 1, \ldots, p, \; c \in \{1, 2\},
\end{equation}
for $q=3$.
By means of testing the hypotheses in \eqref{testing}, we make binary decisions (rejection or non-rejection) for both baseline-category sets, consisting of $p$ null hypotheses each, based on the data at hand. In the MALDI context, this means that we aim at identifying the most distinctive m/z values for the cancer association, where three different cancer types are considered in our application.

\subsection{Multiple Marginal Models}
The following (second) step of the proposed framework is to combine all marginal fits and to approximate the joint null distribution of all ML estimators within each baseline-category pair.
To do so, we exploit the approach for jointly estimating the parameters of multiple marginal models developed by Pipper et al.\cite{mmm}, and apply this approach to the marginal models explained in the previous section. 
Notice that our modelling approach implies that the regression parameters are unique to one model $j$ and are not shared by any two models $j_1 \neq j_2$. In addition to that, the intercepts $(\alpha_{jc})_{1 \leq j \leq p, \thickspace 1 \leq c \leq 2 }$ are nuisance parameters in the sense that the hypotheses in \eqref{testing} solely concern the $\beta_{jc}$'s. However, the intercepts $(\alpha_{jc})_{1 \leq j \leq p,  \thickspace c \in \{1, 2\}}$ contribute to the estimation and the standardization of $\beta_{jc}$'s.

We aim at establishing a central limit theorem for the two $p$-dimensional random vectors $\hat{\beta}_{c} = ( \hat{\beta}_{1c}, \ldots, \hat{\beta}_{pc})_{1 \leq c \leq 2}^\top$. This is accomplished by stacking the standardised score contributions of the $\hat{\beta}_{jc}$'s across all $p$ marginal models, within each baseline-category pair. To this end, we use that, under standard regularity assumptions (like  finite variances and non-vanishing limiting relative category frequencies), each ML estimator  $\hat{\beta}_{jc}$ admits the asymptotic (i.\ e., $n \to \infty$) representation \cite{mmm}
\begin{equation}\label{asym-univ}
(\hat{\beta}_{jc} - \beta_{jc}) \sqrt{n} = \frac{1}{\sqrt{n}} \sum_{i = 1}^{n} \Psi_{ijc} + o_{\mathbb{P}}(1),
\end{equation}
where $\Psi_{ijc} =  \left(F^{(i)}_j\right)^{-1}\widetilde{\Psi}_{ijc}$,
$\left(F^{(i)}_j\right)^{-1}$ is the relevant row of the inverse Fisher information matrix of the model specified in \eqref{logit} that corresponds to coordinate $j$ for the $i$-th observational unit, $\widetilde{\Psi}_{ijc}$ is the score function pertaining to coordinate $j$ and category $c$ for the $i$-th observational unit, and $o_{\mathbb{P}}(1)$ indicates a sequence of random variables converging to zero in probability. 

Elementary calculations yield that $(\Psi_{ij1}, \Psi_{ij2})^\top$ is given by the second and the fourth coordinate of the four-variate vector
\begin{equation}
\label{mmmcontributions}
 \left(\pi_{jm}^{(i)}[\mathbf{I}(m = m')-\pi_{jm'}^{(i)}](1, X^{(i)}_{j})^\top (1, X^{(i)}_{j})\right)_{1 \leq m, m' \leq 2}^{-1} \left(Y^{(i)}_{1} - \pi_{j1}^{(i)}, Y^{(i)}_{2} - \pi_{j2}^{(i)}\right)^\top \otimes (1, X^{(i)}_{j})^\top, \\
\end{equation}
where 
\[\mathbf{I}(m = m') =
  \begin{cases}1,~&{\text{ if }}~ m = m',\\0,~&{\text{ if }}~  m \neq m', \end{cases}\]
\[{\pi}_{jm}^{(i)} = \frac{\exp\left(\alpha_{jm} + X_{j}^{(i)}\beta_{jm}\right)}{1 + \displaystyle \sum_{h =1}^{2}\exp\left(\alpha_{jh} + X_{j}^{(i)}\beta_{jh}\right)},
\]
and $\otimes$ denotes the Kronecker product. Moreover, the Fisher information matrix of the full sample of size $n$ is the sum of the Fisher information matrices pertaining to each $i \in \{1, \ldots, n\}$, due to the assumption of stochastically independent observational units. For more (computational) details regarding multinomial regression, see, for example, \cite{boehning, mvt}.

Now, let $\beta_{c} := ({\beta}_{1c}, \ldots, \beta_{pc})^\top$, 
$\hat{\beta}_{c} := (\hat{\beta}_{1c}, \ldots, \hat{\beta}_{pc})^\top$, and $\Psi_{ic} := (\Psi_{i1c}, \ldots, \Psi_{ipc})^\top$ for $c \in \{1, 2\}$ and $i \in \{1, \ldots, n\}$. Then, we can conclude from \eqref{asym-univ} the asymptotic expansion
\begin{equation}
\label{asym}
(\hat{\beta_{c}} - \beta_{c}) \sqrt{n} = \frac{1}{\sqrt{n}} \sum_{i = 1}^{n} \Psi_{ic} + o_{\mathbb{P}}(1),
\end{equation}
where now $o_{\mathbb{P}}(1)$ indicates a sequence of $p$-dimensional random vectors converging to the zero vector in probability. Since the observational units are assumed to be stochastically independent and identically distributed, the multivariate central limit theorem yields convergence in distribution (indicated by the symbol $\xrightarrow{d}$) to a centered $p$-variate normal distribution, i.\ e., that
\begin{equation}
(\hat{\beta_{c}} - \beta_{c}) \sqrt{n} \xrightarrow{d} N_p(0, \Sigma_{c}) \text{~~for~~} c \in \{1, 2\}.
\end{equation}
For each $c \in \{1, 2\}$, the limiting variance-covariance matrix $\Sigma_{c}$ can be estimated consistently by
\begin{equation}
\label{mmmvariance}
\widehat{\Sigma}_{c} = \frac{1}{n} \sum_{i = 1}^{n} \hat{\Psi}_{ic}^{\top}\hat{\Psi}_{ic},
\end{equation}
where each $\hat{\Psi}_{ic}$ is obtained by inserting the parameter estimates $(\hat{\alpha}_{jm}, \hat{\beta}_{jm})_{1 \leq j \leq p, \; m \in \{1, 2\}}$ from the $p$ marginal fits (cf.\ \eqref{mle}) instead of the true parameter values $(\alpha_{jm}, \beta_{jm})_{1 \leq j \leq p, \; m \in \{1, 2\}}$ into $\Psi_{ic}$.

For each $c \in \{1, 2\}$, let $Z_{1c}, \ldots, Z_{pc}$ be the standardised versions of $\hat{\beta}_{1c}, \ldots \hat{\beta}_{pc}$, which are given by
\begin{equation}
\label{z_values}
Z_{jc} = \frac{\hat{\beta}_{jc}}{\sqrt{\widehat{\text{Var}}({\hat{\beta}_{jc}})}}, \quad j = 1, \ldots, p, \quad  c \in \{1, 2\},  \\
\end{equation}
where $\sqrt{\widehat{\text{Var}}(\hat{\beta}_{jc})}$ is the square root of the $j$-th diagonal element of $\hat{\Sigma}_{c}$, divided by  $\sqrt{n}$. We have that
\begin{equation}
\label{zstat}
(Z_{1c}, Z_{2c}, \ldots, Z_{pc})^{\top} \underset{\text{approx.}}{\sim} N_p((\mu_{1c}, \mu_{2c}, \ldots, \mu_{pc})^{\top}, \hat{\Sigma}_{c}^{*}),
\end{equation}
where $\mu_{jc} = \beta_{jc} / \sqrt{\widehat{\text{Var}}(\hat{\beta}_{jc})}$ for $1\leq j \leq p$ and $ 1\leq c \leq 2$,  $\hat{\Sigma}_{c}^* = \diag[\hat{\Sigma}_{c}]^{-1/2} \hat{\Sigma}_{c} \diag[\hat{\Sigma}_{c}]^{-1/2}$ denotes the correlation matrix pertaining to $\hat{\Sigma}_{c}$ from \eqref{mmmvariance}, and the notation $\underset{\text{approx.}}{\sim}$ indicates the approximate distribution for large $n$. Therefore, since $\sqrt{\widehat{\text{Var}}(\hat{\beta}_{jc})} > 0$ holds true with probability one, the families of hypotheses from \eqref{testing} can be equivalently written as  
\begin{equation}\label{test}
{H_{0jc}: \mu_{jc} = 0 } \text{~~versus~~} {H_{1jc}: \mu_{jc} \neq 0 }, \;\; j = 1, \ldots, p, \; c \in \{1, 2\}.
\end{equation}
For the sake of presentation, we use henceforth $\hat{\Sigma}_{c}^{*}$ and $Z_{1c}, \ldots, Z_{pc}$ only for the first baseline-category pair (i.\ e., for $c = 1$) and omit the index $c$, for convenience of notation. However, we consider the same inference procedure for both baseline-category pairs.
\subsection{False Discovery Proportion}
\label{section:fdp}
Considering the multiple test problem given by \eqref{test} (for $c=1$), let $p_{0} = \vert\{j: \mu_{j} = 0\}\vert$ denote the number of true null hypotheses, and $p_{1} = p - p_0 = \vert\{j: \mu_{j} \neq 0\}\vert$ the number of false null hypotheses. For controlling type I errors, we employ empirical process techniques, in the spirit of Storey's method.\cite{storey2002}
In particular, we consider thresholding rules which reject a null hypothesis $H_{0j}$ if and only if a corresponding $p$-value $p_j$ is smaller than or equal to a (data-dependent) threshold ${t}$. Such types of multiple tests have been considered widely in previous literature. \cite{pfa, pfa_poet, storey2002}
 Conceptually, the goal of the proposed method is to approximate the proportion of false discoveries among all rejections (commonly referred to as the false discovery proportion, FDP for short) for a fixed threshold $t$ under an arbitrary correlation matrix $\Sigma^{*}$.
To this end, we consider the three empirical processes 
\begin{align*}
V(t) &= \#\{\text{true null~} P_{j}: P_{j} \leq t\},\\
S(t) &= \#\{\text{false null~} P_{j}: P_{j} \leq t\},\\
R(t) &= \#\{P_{j}: P_{j} \leq t\},
\end{align*}
where ${t}$ ranges in $[0, 1]$ and the notation $P_j$ is used to indicate that we consider the $j$-th $p$-value as a random variable here. 
For fixed $t \in [0, 1]$, the quantities $V(t)$, $S(t)$, and $R(t)$ represent
the (random) number of false discoveries (i.\ e., false rejections or, synonymously, type I errors), the (random) number of true discoveries, and the (random) total number of discoveries, respectively, hence $R(t) = V(t) + S(t)$. The FDP is for a fixed $t$ given by $\text{FDP}(t) = V(t) / \max\{R(t), 1\}$, where the maximum is taken to avoid an expression of the form $0/0$.
All of the latter random variables depend on the $Z$-statistics $Z_{1}, Z_{2}, \ldots, Z_{p}$, because each (random) $p$-value $P_j$ will be obtained by a transformation of $Z_j$, for $1 \leq j \leq p$, see below. 
 Notice also that $V(t)$ and $S(t)$ are both unobservable, whereas $R(t)$ is observable. 
\subsection{Principal Factor Approximation}
\label{section:pfa}
The next step of our proposed framework is to incorporate the correlation effects of the 
$Z$-statistics in an approximation of $\text{FDP}(t)$ for a fixed threshold $t$. The procedure relies on the identification of a low-dimensional linear space of random vectors which captures most of the dependence structure of the $Z$-statistics. 

To this end, we carry out the spectral decomposition of the estimated correlation matrix $\hat{\Sigma}^{*}$. We let $\lambda_{1}, \ldots ,\lambda_{p} \geq 0$ denote the eigenvalues of $\hat{\Sigma}^{*}$, and $\gamma_{1}, \ldots, \gamma_{p}$ the corresponding eigenvectors. Hence, $\hat{\Sigma}^*$ is represented by its eigenvalue-eigenvector pairs $(\lambda_j, \gamma_j)_{1 \leq j \leq p}$. For a fixed number $k$ of common factors, where $1 \leq k << p$, the factor approximation of $\hat{\Sigma}^{*}$ is given by
 \begin{equation}
 \hat{\Sigma}^* = L_k L_k^{\top} + A_k,
 \end{equation}
 where $A_k = \sum_{j = k + 1}^{p} \lambda_{j}\gamma_{j}\gamma_{j}^{\top}$ and  $L_k = (\sqrt{\lambda_{1}}\gamma_{1}, \sqrt{\lambda_{2}}\gamma_{2}, \ldots,  \sqrt{\lambda_{k}}\gamma_{k})$. 
 If $k$ is chosen appropriately, the $p \times k$- matrix $L_k$, corresponding to $k$ latent variables, can describe most of the dependence structure among the $Z$-statistics. 
Accordingly, $Z_{1}, \ldots, Z_{p}$ can be decomposed as
\begin{equation}
\label{pfa_eq}
Z_{j} = \mu_{j} + b_{j}^{\top} W + K_{j} = \mu_{j} + \eta_j + K_{j}, \quad j = 1, \ldots, p,
\end{equation}
where $b_{j} = (b_{j1}, \ldots, b_{jk})^{\top}$ and  $(b_{1j}, \ldots, b_{pj})^{\top} = \sqrt{\lambda_{j}}\gamma_{j}$. The vector $W = (W_{1}, \ldots, W_{k})^\top \sim N_k(0, I_k)$ is the vector of (latent) common factors. Due to orthogonality projection, these common factors are jointly stochastically independent. The random vector $(K_{1}, \ldots ,K_{p})^\top \sim N_p(0, A_k)$ is called the vector of random errors, and it is assumed that factors and random errors are stochastically independent. In our previous work \cite{vutovdickhaus}, we elaborated more on the importance and the choice of $k$.

On the basis of the above considerations, we consider the FDP estimator from Proposition 2 by Fan et al. \cite{pfa}, which is for a given $t \in [0, 1]$ defined as
 \begin{equation}
 \label{pfa_est}
 \widehat{\text{FDP}}(t) = \min \left\{ \sum_{j =  1 }^{p} \left[\Phi(a_{j}(z_{t/2} + \widehat{\eta}_{j}) + \Phi(a_{j}(z_{t/2} - \widehat{\eta}_{j}))\right], R(t)\right\} / R(t)
 \end{equation}
 if $R(t) >  0$, and  $\widehat{\text{FDP}}(t) = 0$ if $R(t) = 0$. In \eqref{pfa_est}, $a_j = (1 - \sum_{h = 1}^{k} b_{jh}^2)^{-1/2}$ and $R(t) = \{j: P_j \leq t \}$, where
 $\Phi$ and $z_{t/2} = \Phi^{-1}(t/2)$ are the cumulative distribution function (cdf) and the lower $t/2$-quantile of the standard normal distribution on $\mathbb{R}$, respectively. 
In this, the (unadjusted) two-sided, random $p$-value corresponding to $Z_j$ is given by $P_j = 2\Phi(-|Z_{j}|) = 2(1 - \Phi(|Z_{j}|))$, and  $\hat{\eta}_{j} = \sum_{h = 1}^{k} b_{jh}\widehat{W}_{j}$ is a linear estimator for $\eta_{j} = b_{j}^\top W$.

The FDP estimator given in \eqref{pfa_est} is based on a sparsity assumption. \cite{pfa, pfa_poet} Specifically, it is assumed that the numbers $p$ and $p_{0}$ are large, while the number $p_{1}$ of false nulls is relatively small. This assumption allows for the summation over all hypotheses (not only over "true nulls") in the numerator of \eqref{pfa_est}. In the context of MALDI modelling, this sparsity assumption is justified, for three reasons: First, several researchers who have analysed MALDI datasets have reported a low number of m/z values that are highly associative for a cancer subtype. \cite{behrmann, boskamp} Second, to the best of our knowledge, the reported number of biologically meaningful molecules, so-called biomarkers, in MALDI-related studies, is low \textendash \, from three to five per study. \cite{krieg2, johannes} Third, there is evidence of a small signal-to-noise ratio \cite{johannes, boskamp}. 

To evaluate \eqref{pfa_est} in a practical application, one needs to determine the (linear) estimator $\widehat{W} = (\widehat{W}_{1}, \ldots, \widehat{W}_{k})^\top$ of the common factors. 
The authors of \cite{pfa} have proposed to estimate $\widehat{W}$ via
a linear regression ($L_2$-regression) or by using a quantile regression ($L_1$-regression).  Regarding the $L_2$ estimator, it has been proposed to carry out the estimation of $(\widehat{W}_{1}, \ldots, \widehat{W}_{k})^\top$ only on the basis of the subset of length $0.95p$ of the smallest (in absolute values) $Z$-statistics. 
This leads to the $L_2$ estimator 
\begin{equation}
\label{ols}
\widehat{W} = (\widehat{W}_1, \ldots, \widehat{W}_k)^\top = \arg \min_{W \in \mathbb{R}^k} \sum_{ j = 1}^{ \lfloor 0.95 p \rfloor } (Z_{j} - b_{j}^\top W)^{2},
\end{equation}
where the Z-values in \eqref{ols} are sorted in ascending order based on their absolute values.
We have utilized this estimator in our simulation study (see Section \ref{sec4}). The estimator based on the $L_1$-regression is given by
\begin{equation}
\label{l1}
\widehat{W} = (\widehat{W}_1, \ldots, \widehat{W}_k)^\top = \arg \min_{W \in \mathbb{R}^k} \sum_{ j = 1}^{p} |Z_{j} - b_{j}^{\top}W|.
\end{equation}
For the analysis of the MALDI data (see Section \ref{sec-MALDI-data}), we have used the $L_1$-regression rather than the $L_2$-regression, because it is more robust to highly untypical observations (i.\ e.,  outliers).

Finally, the dependency-adjusted (random) $p$-values pertaining to $Z_{1}, \ldots, Z_{p}$ are given by  
\begin{equation}
\label{adjmethod}
\tilde{P}_{j} = 2\Phi(-|a_{j}(Z_{j} - b_{j}^{\top}\widehat{W})|).
\end{equation}
For a given threshold $t \in [0, 1]$, the null hypothesis $H_{0j}$ from \eqref{test} gets rejected based on the data at hand, if and only if $\tilde{p}_{j} \leq {t}$, $1 \leq j \leq p$. For the purpose of FDP control, $t$ can be selected as the largest value $t = t_{\alpha} \in [0, 1]$ fulfilling that $\widehat{\text{FDP}}(t_{\alpha})$ is not exceeding a pre-determined level $\alpha \in (0, 1)$, for example, $\alpha = 10\%$. In practice, one can carry out a grid search over a set of candidate values of $t$ in order to find the value $t_{\alpha}$ for a given $\alpha$. \cite{vutovdickhaus}

\section{Simulation Study}\label{sec4}
\subsection{Simulation Setup}
  We carried out a simulation study to assess the performance of our proposed methodology under different data-generating schemes. We considered the parameter settings $n = 500$, $p \in \{ 500, 1000 \}$, and $p_{1} = 10$ (for each baseline-category pair).  For each combination of the aforementioned parameter values, we performed $1,000$ Monte Carlo repetitions. Without loss of generality, we set $\beta_{j1} \neq 0$  as well as $\beta_{j2} \neq 0$ for $j \in \{1, \ldots, p_{1}\}$, and we call these first $p_1$ coordinates "active". The remaining $p_0$ coordinates have been set "inactive", meaning that $\beta_{jc} = 0$ for all in $j \in \{p_1 +1, \ldots, p\}$ and $c \in \{1,2 \}$. Furthermore, we set all intercept terms $(\alpha_{jc})_{1 \leq j \leq p,\; c \in \{1, 2\}}$ to zero in all simulations. We estimated the common factors $\{W_{h} : h = 1, \ldots, k\}$ by applying the least-squares estimator as given in \eqref{ols}. The utilized number $k$ of common factors is reported in each caption of Tables \ref{table:sim_1_1} - \ref{table:sim_2_2} below.
  
   For each observational unit $i \in \{1, \ldots, n\}$, the simulation data have been generated according to the model 
   \[\mathbb{P}_\beta(Y^{(i)}_{1} = 1 | X^{(i)} ) = \frac{\exp\left(\sum_{j=1}^{p_1} \beta_{j1} X_{j}^{(i)}\right)}{1 + \exp\left(\sum_{j=1}^{p_1} \beta_{j1} X_{j}^{(i)}\right) + \exp\left(\sum_{j = 1}^{p_{1}} \beta_{j2} X_{j}^{(i)}\right)},
   \]
  \[\mathbb{P}_\beta(Y^{(i)}_{2} = 1 | X^{(i)} ) = \frac{\exp\left(\sum_{j=1}^{p_1} \beta_{j2} X_{j}^{(i)}\right)}{1 + \exp\left(\sum_{j=1}^{p_1} \beta_{j1} X_{j}^{(i)}\right) + \exp\left(\sum_{j = 1 }^{p_{1}} \beta_{j2} X_{j}^{(i)}\right)}.
   \]
   The considered dependency structures for $X_1, \ldots, X_p$ are given in Model \ref{model-simu}.
   \begin{model}\label{model-simu} $ $\\
   \onehalfspacing\underline{Scenario 1:} $X_{1}, \ldots, X_{p}$ are stochastically independent and identically standard normally distributed random variables. \\
   \onehalfspacing\underline{Scenario 2:} 
   $X_{p_{1}+1}, \ldots, X_{p}$ are jointly normally distributed on $\mathbb{R}^{p_0} $, namely, according to $N_{p_0}(0, \Sigma^{*})$, where $\Sigma^{*}$ is an equi-correlation matrix with diagonal elements equal to one, and off-diagonal elements equal to some given value $\rho$. Furthermore, the subvector $(X_{p_{1}+1}, \ldots, X_{p})^\top$ is stochastically independent of the
   subvector $(X_1, \ldots, X_{p_1})^\top$. This is in order to avoid spurious effects of covariates $X_j$ with $p_{1} +1 \leq j \leq p$ on the response, which would arise from confounding if the two aforementioned subvectors would be dependent. The random variables $X_1, \ldots, X_{p_1}$ are stochastically independent and identically $N(0, 1)$-distributed. The equi-correlation model is a one-factor model with a dominating first eigenvalue; see, e.\ g., Example 2.1 in \cite{FiDiRo2007}.
    \end{model}

This simulation study has been run in \textit{R version 4.1.1}, \cite{rbase} employing the R function \textit{rmvnorm()} \cite{mvtnorm} to simulate correlation dependency among the independent variables. In addition, we have employed the function \textit{vglm()} \cite{vgam} for the estimation of the marginal multinomial estimates. The function \textit{pfa.test()} \cite{r_pfa} has been used for applying the PFA method.

\subsection{Simulation Results}
We summarize the results from our computer simulations in Tables \ref{table:sim_1_1} -  \ref{table:sim_2_2} in terms of $\widehat{\text{FDP}}(t)$, $R(t)$, and $S(t)$ for a fixed threshold $t = 10^{-4}$. In addition, we report the median value of $t_{\alpha}$ for the conventional choice of $\alpha = 0.05$ as well as the average (over the $1{,}000$ Monte Carlo repetitions) of $S(t_{\alpha})$. Due to the choice of $p_1 = 10$, the largest possible value of $S(t)$ equals ten, for any choice of the threshold $t$.

Tables \ref{table:sim_1_1} and \ref{table:sim_1_2} contain our simulation results under Scenario 1 from Model \ref{model-simu}. Because the $Z$-statistics are jointly stochastically independent (due to joint independence among all $X_j$), $t_{0.05}$ is rather small here, because the "effective number of tests" (in the sense of Section 3.4 in \cite{handbook-chapter}) equals $p$ under joint independence of all test statistics or $p$-values, respectively. This means that a rather strong multiplicity correction is necessary. However, the standard error of $\widehat{\text{FDP}}(t)$ is also quite small under Scenario 1, since the FDP is well concentrated around its expectation (the FDR) under joint independence among all test statistics or $p$-values, respectively. With regard to type II errors, the reported average values of $S(t_{0.05})$ demonstrate that on average $7 - 8$ out of $10$ active coordinates could be identified by our proposed multiple test procedure.

Simulation results under Scenario 2 (Gaussian equi-correlation model) are summarised in Tables \ref{table:sim_2_1} and  \ref{table:sim_2_2}. Here,  the effective number of tests is smaller than $p$ whenever $\rho \ge 0$, and it decreases as $\rho$ increases. Consequently, $t_{0.05}$ grows with $\rho$, too. In turn, this also leads to an improved power of the proposed multiple test, which is reflected by the numbers reported for $S(t)$, which are under Scenario 2 on average larger than the corresponding values under Scenario 1. 
\begin{table}[ht]
		\caption{\bf Simulation results under Scenario 1 (I)}
		\begin{tabular}{@{\extracolsep{0.1pt}}lccccccccc}
			\hline \\
			${c}$ & \shortstack{Median of\\ $\widehat{\text{FDP}}(t)$} & \shortstack{ Std. Error \\ of $\widehat{\text{FDP}}(t)$}   &\shortstack{Mean of\\ $R(t)$} & \shortstack{Std. Error \\ of $R(t)$} & \shortstack{Mean of \\ $S(t)$} & \shortstack{Std. Error \\ of $S(t)$} & \shortstack{Median of $t_{0.05}$} & \shortstack{Mean \\ $S(t_{0.05})$} \\
			\hline\\
		$1$& 0.00433 & 0.00147  & 5.80  &  1.39 & 5.75 &  1.36 &  1.60e-03 & 8.37\\ 
			\hline \\
		$2$ &	0.00442 &  0.00134 & 5.78 & 1.36 & 5.74  & 1.35 & 1.60e-03 & 8.31 \\
			\hline
		\end{tabular}
			\begin{flushleft}The total number of hypotheses equals $p = 500$; the number of false null hypotheses equals $p_{1} = 10$ per baseline-category pair (non-zero regression coefficients equal $1$); the number of factors equals $k = 10$; the rejection threshold equals $t = 10^{-4}$, apart from the last two columns.
	\end{flushleft}
		\label{table:sim_1_1}
\end{table}

\begin{table}[ht]
		\caption{\bf Simulation results under Scenario 1 (II)}
		\begin{tabular}{@{\extracolsep{0.1pt}}lcccccccccc}
			\hline \\
			${c}$ &\shortstack{Median of\\ $\widehat{\text{FDP}}(t)$} & \shortstack{ Std. Error \\ of $\widehat{\text{FDP}}(t)$}   &\shortstack{Mean of\\ $R(t)$} & \shortstack{Std. Error \\ of $R(t)$} & \shortstack{Mean of \\ $S(t)$} & \shortstack{Std. Error \\ of $S(t)$} & \shortstack{Median of $t_{0.05}$} & \shortstack{Mean \\ $S(t_{0.05})$} \\
			\hline\\
		$1$ &	0.00946 & 0.00325  & 5.94  &  1.35 & 5.86 &  1.33 &  5.52e-04 & 7.50\\  
			\hline \\
		$2$ &	0.00943 &  0.00360 & 5.83 & 1.43 & 5.76  & 1.40 & 5.52e-04 & 7.46 \\ 
			\hline
		\end{tabular}
			\begin{flushleft}The total number of hypotheses equals $p = 1000$; the number of false null hypotheses equals $p_{1} = 10$ per baseline-category pair (non-zero regression coefficients equal $1$); the number of factors equals $k = 10$; the rejection threshold equals $t = 10^{-4}$, apart from the last two columns.
	\end{flushleft}
		\label{table:sim_1_2}
\end{table}
\begin{table}[ht]
		\caption{\bf Simulation results under Scenario 2 (I)}
		\begin{tabular}{@{\extracolsep{0.1pt}}lccccccccccc}
			\hline \\
			${c }$ & $\rho$ &  \shortstack{Median of\\ $\widehat{\text{FDP}}(t)$} & \shortstack{ Std. Error \\ of $\widehat{\text{FDP}}(t)$}   &\shortstack{Mean of\\ $R(t)$} & \shortstack{Std. Error \\ of $R(t)$} & \shortstack{Mean of \\ $S(t)$} & \shortstack{Std. Error \\ of $S(t)$} & \shortstack{Median of $t_{0.05}$} & \shortstack{Mean \\ $S(t_{0.05})$} \\
			\hline\\
		$1$ &		0.2 & 0.00287 & 0.0199  & 5.83  &  1.36 & 5.81 &  1.31 &  2.33e-03 & 8.63\\
			\hline \\
		$2$ &		0.2 & 0.00273 &   0.00995 & 5.90 & 1.33 & 5.89  & 1.31 & 2.33e-03 & 8.72 \\
			\hline \\
	$1$ &		0.5 & 0.000274 & 0.0475  & 5.84  & 1.48 & 5.81 &  1.31 &  6.99e-03 & 9.28\\  
			\hline \\
	$2$ &	0.5 & 0.000265 &  0.0168 & 5.88 & 1.33 & 5.89  & 1.31 & 6.99e-03 & 9.31 \\ 
			\hline \\
	$1$ &		0.8 & 0.000171 & 0.0455  & 5.87  &  2.46 & 5.81 &  1.31 &  2e-02 & 9.77\\  
			\hline \\
$2$ &	  0.8 & 0.000168 &  0.0141 & 5.86 & 1.31 & 5.90  & 1.32 & 2e-02 & 9.78 \\ 
			\hline
		\end{tabular}
			\begin{flushleft}The total number of hypotheses equals $p = 500$; the number of false null hypotheses equals $p_{1} = 10$ per baseline-category pair (non-zero regression coefficients equal $1$); the number of factors equals $k = 1$; the rejection threshold equals $t = 10^{-4}$, apart from the last two columns.
	\end{flushleft}
		\label{table:sim_2_1}
\end{table}

\begin{table}[ht]
		\caption{\bf Simulation results under Scenario 2 (II)}
		\begin{tabular}{@{\extracolsep{0.1pt}}lccccccccccc}
			\hline \\
		${c }$ &	$\rho$ &  \shortstack{Median of\\ $\widehat{\text{FDP}}(t)$} & \shortstack{ Std. Error \\ of $\widehat{\text{FDP}}(t)$}   &\shortstack{Mean of\\ $R(t)$} & \shortstack{Std. Error \\ of $R(t)$} & \shortstack{Mean of \\ $S(t)$} & \shortstack{Std. Error \\ of $S(t)$} & \shortstack{Median of $t_{0.05}$} & \shortstack{Mean \\ $S(t_{0.05})$} \\
			\hline\\
		$1$ &		0.2 & 0.00497 & 0.0208  & 5.86  &  1.36 & 5.81 &  1.32 &  1.01e-03 & 8.00 \\
			\hline \\
	$2$ &	0.2 & 0.00505 &   0.0265 & 5.84 & 1.38 & 5.79  & 1.32 & 1.01e-03 & 7.98 \\ 
			\hline \\
	$1$ &	0.5 & 0.000299 & 0.0316 & 5.84 &  1.37 & 5.81 &  1.32 &  5.06e-03 & 9.08 \\  
			\hline \\
	$2$ &	0.5 & 0.000315 &  0.0496 & 5.84 &  1.63 & 5.79  & 1.32 & 5.06e-03 & 9.07 \\ 
			\hline \\
$1$ &	0.8 & 0.000168 & 0.0122  & 5.81  &  1.33 & 5.81 &  1.32 &  2.9e-02 & 9.73\\  
			\hline \\
	 $2$ &      0.8 & 0.000169 &  0.0402 & 5.84 & 2.68 & 5.79  & 1.32 & 2.9e-02 & 9.75 \\ 
			\hline
		\end{tabular}
			\begin{flushleft}The total number of hypotheses equals $p = 1000$; the number of false null hypotheses equals $p_{1} = 10$ per baseline-category pair (non-zero regression coefficients equal $1$); the number of factors equals $k = 1$; the rejection threshold equals $t = 10^{-4}$, apart from the last two columns.
	\end{flushleft}
		\label{table:sim_2_2}
\end{table}
\section{Real Data Application: MALDI Imaging Data}\label{sec-MALDI-data}
\subsection{Description of the Dataset}
We applied our proposed inferential procedure to a hyperspectral dataset obtained from a MALDI imaging instrument. For a detailed description of this dataset in terms of its acquisition protocols, tissue sections, tissue blocks, etc., see \cite{behrmann, boskamp}.
In the aforementioned studies, the researchers analysed this dataset by merging two lung cancer subtypes into a single category. By doing so, the authors worked with two cancer subtypes based on two distinctive human organs (the so-called "LP task", i.e. lung vs. pancreas). In our study, however, we model this dataset with a three-class model, using the subtype of pancreatic adenocarcinoma as the baseline. The resulting goal of the statistical analysis is to perform simultaneously two (sub-) tasks, namely the pancreatic vs. lung adenocarcinoma and the pancreactic vs. lung squamous cell carcinoma comparison.

\begin{figure}[h]
  \includegraphics[height = 0.60\textwidth, width=0.95\textwidth]{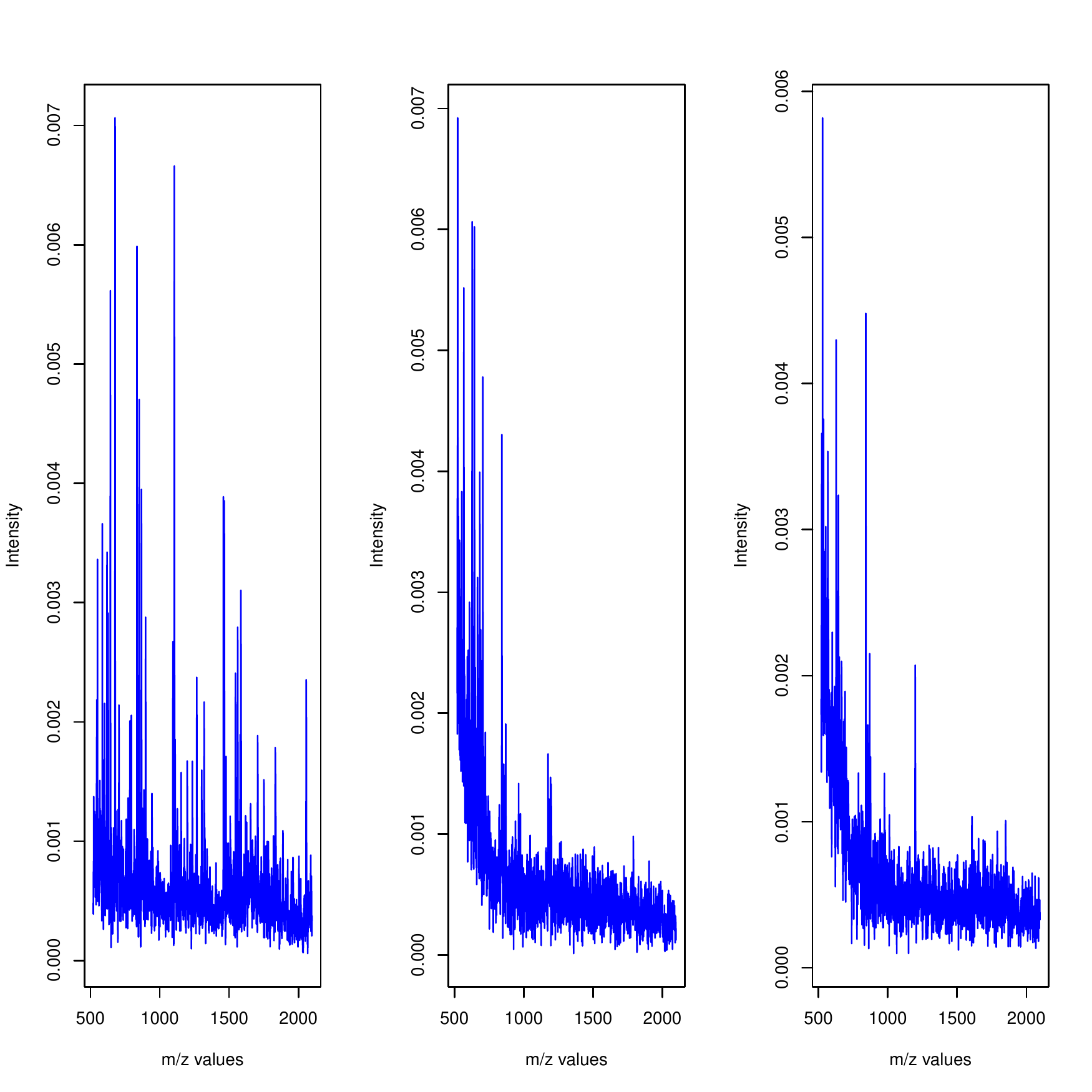}
  \caption{Examples of mass spectra for three cancer subtypes.  }
  \label{fig:mz_values}
\end{figure}

As mentioned in the introduction, the high-throughput device that motivated this study is MALDI IMS. In general, this technology provides  molecular information about a given analyte (for example, tissue) in a spatial manner. More concretely, the MALDI IMS tool measures mass spectra at multiple discrete spatial positions and yields an image for each unique spot within a provided tissue.  For an illustration, Figure \ref{fig:mz_values} displays instances of outputs from a MALDI experiment from three spatial spots. In Figure \ref{fig:mz_values}, the m/z values are plotted on the horizontal axes, and the relative abundances (intensities values) of ionizable molecules are plotted on the vertical axes. This "spatial molecular (biochemical) information can then be used for the determination of the cancer subtypes or the identification of the origin of the primary tumour in patients"\cite{behrmann}. Each spot is called a mass spectrum\cite{alex, alex2}, and it depicts relative abundances of ionizable molecules with numerous mass-to-charge (m/z) values \textendash \, "ranging from several hundred up to a few tens of thousands m/z values" \cite{alex2} \textendash \, whereas a single m/z value in the context of MALDI is interpreted as a molecular mass. 
The full pipeline of a MALDI experiment, from a tissue spot to data analysis, is thoroughly illustrated in Figure 1 of \cite{boskamp} and Figure C2 of \cite{johannes}. In the dataset analysed here, formalin-fixed paraffin-embedded (FFPE) \cite{samplepreparation} tumour samples from 445 patients were provided by the tissue bank of the National Center for Tumor Diseseas (NCT) Heidelberg, Germany. Tissue cores of three cancer subtypes were used, comprising lung adenocarcinoma (ADC) (168 patients), lung squamous cell carcinoma (SqCC) (136 patients) and pancreatic adenocarnimoma (141 patients). 
The corresponding dataset is publicly avaible from (\url{https://gitlab.informatik.uni-bremen.de/digipath/Deep_Learning_for_Tumor_Classification_in_IMS} \textendash \, last accessed 7 February 2022). 

\subsection{Data Preprocessing}
We performed the following (further) data preprocessing steps. First, due to the high dimensionality of the feature space ($27,286$ m/z values), spectral filtering was carried out. Namely, we loaded the data into MATLAB R2018, and used the internal library MSClassifyLib (function MSAdaptiveResampleTrigger()) to apply the spectral filtering with default value of 0.4 Da (for more details, see, e.g., \cite{johannes, boskamp} and references therein). Second, the mass range was pruned up to the mass range of 2100 m/z, and we used one mass spectrum from each patient. Finally, the well-established normalisation step TIC (total ion count) \cite{alex} was employed on the chosen mass spectra range. The result of this data preprocessing was a dataset with 445 mass spectra (observational units) and 1579 m/z channels (features).
\subsection{Analysis of the MALDI Dataset}
We have modelled each pixel (based on the averaging filter around 0.4 Da as explained in the previous section) individually to determine those m/z values that are highly associative for a certain cancer subtype. Thus, the m/z values take the role of the $X_j$'s and the cancer subtype takes the role of $Y$ from our general setup, where $j \in \{1, \ldots, p = 1{,}579\}$ here. Statistically speaking, our procedure identifies significant differences in distributions across the subtypes, and is not only identifying significant peaks in the spectra.

\begin{figure}[ht]
    	\begin{minipage}[c]{0.50\linewidth}
    		\subcaption{Task: Pancreatic vs. ADC}
    		\includegraphics[height = 1\textwidth, width=0.9\textwidth]{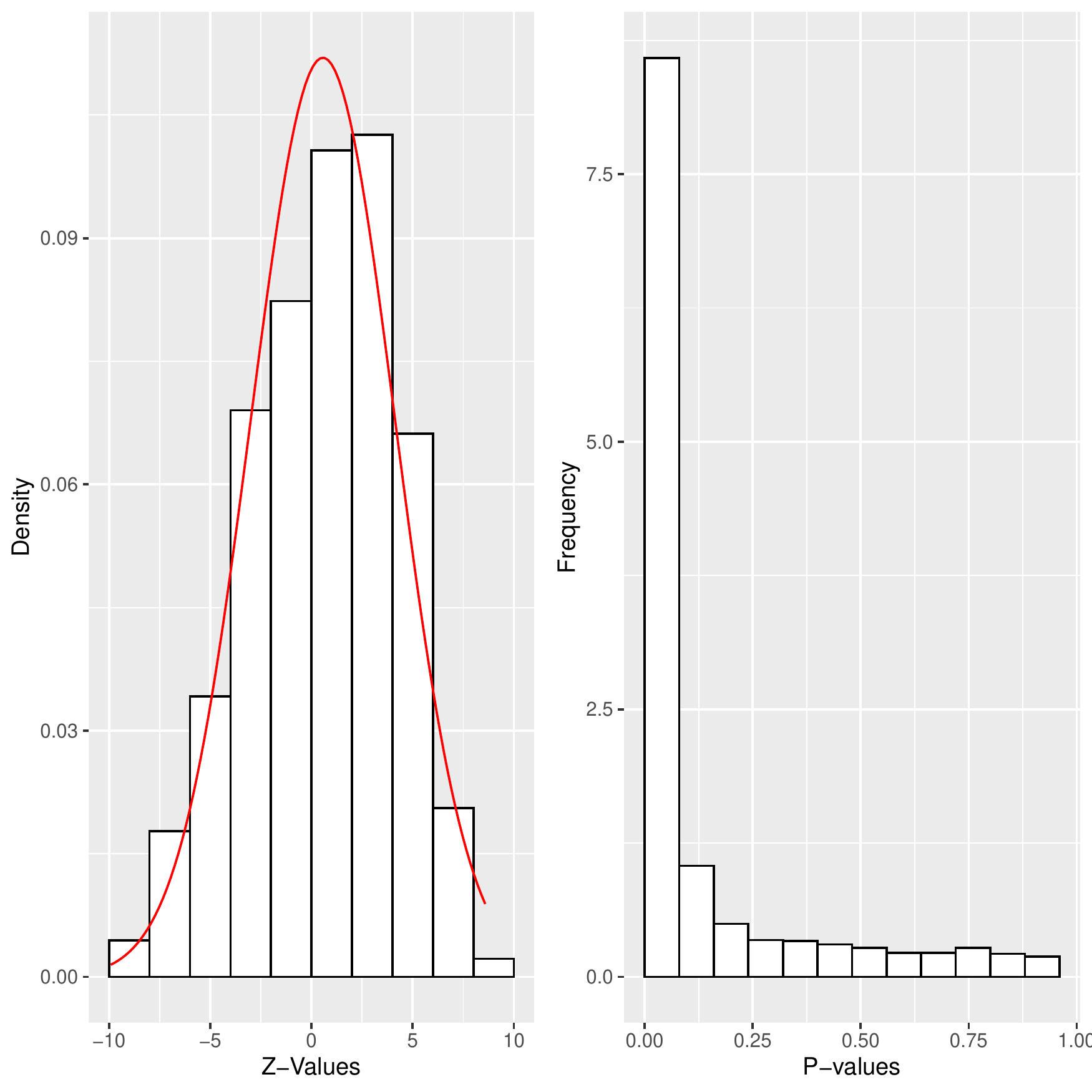}
    	\end{minipage}
    	\hfill
    	\begin{minipage}[c]{0.50\linewidth}
    		\subcaption{Task: Pancreatic vs. SqCC}
    		\includegraphics[height = 1\textwidth, width=0.9\textwidth]{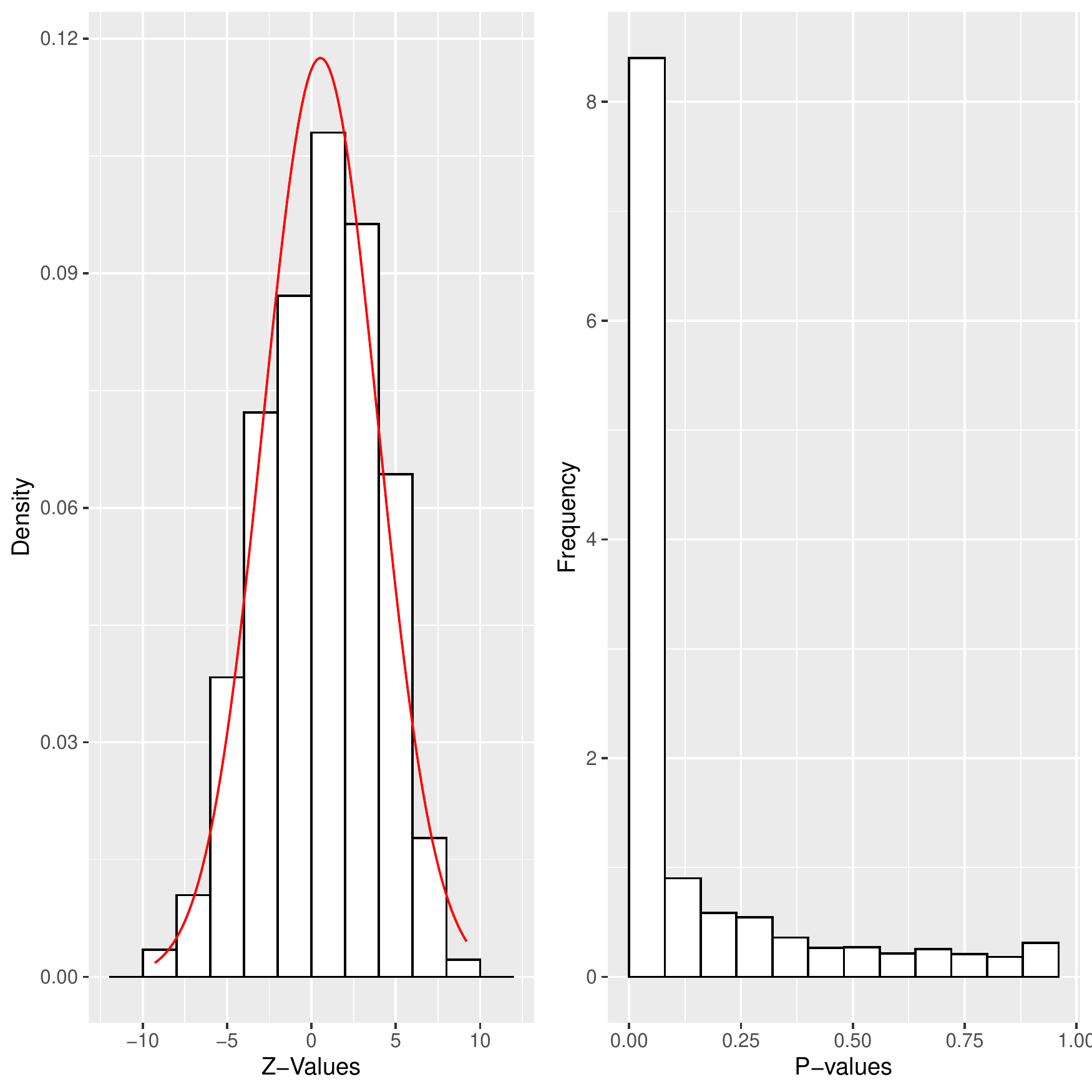}
    	\end{minipage}%
    		\caption{Empirical distributions and fitted normal density curves of the Z-values, as well as histograms of the unadjusted $p$-values, for both tasks. The fitted normal distributions are $N(0.565, 3.365^2)$ for the first task and  $N(0.543, 3.394^2)$ for the second 
    		task. As a result of overdispersion and non-zero effect sizes, there are many non-adjusted $p$-values which are close to zero, in both graphs.}
    		\label{fig:emp_z_values}
    \end{figure}

Figure \ref{fig:emp_z_values} displays the empirical distributions of the Z-statistics, along with histograms of the non-adjusted $p$-values, for both tasks. Clearly, the $Z$-values (for both tasks) do not resemble the realization of a random sample from the theoretical null distribution, which is the standard normal distribution on $\mathbb{R}$. In particular, the empirical variances are much larger than one in both histograms displayed in Figure \ref{fig:emp_z_values}. There are two reasons for this overdispersion, namely, (i) the presence of correlation effects among the $Z$-values (for more details, see \cite{efron2007, efron2010}) and (ii) the presence of extremely large (in absolute value) $Z$-values, which is likely caused by significant effects, peaks and isotopic patterns. In our previous study\cite{vutovdickhaus}, we have observed a similar phenomenon and have described a procedure (based on the "empirical null distribution") which can be used to account for the part of the overdispersion which is caused by correlations among the $Z$-statistics.

\begin{figure}
     \centering
     \begin{subfigure}[ht]{\textwidth}
         \centering
        \caption{Task: Pancreatic vs. ADC.}
        \includegraphics[height = 0.5\textwidth, width=0.9\textwidth]{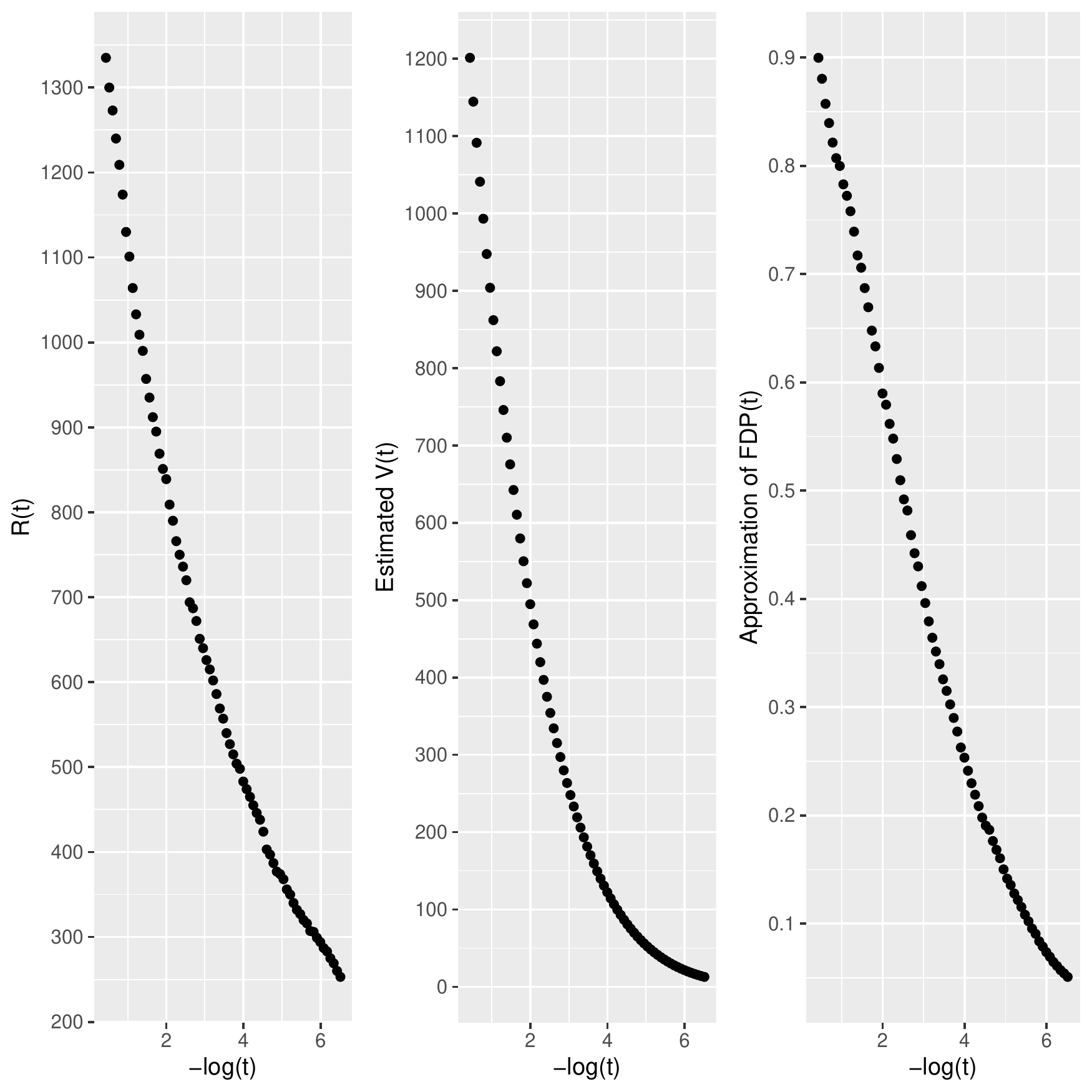}
          \label{fig:ad_panc}
     \end{subfigure}
     \hfill
     \hfill
     \begin{subfigure}[ht]{\textwidth}
         \centering
            \caption{Task: Pancreatic vs. SqCC.}
         \includegraphics[height = 0.5\textwidth, width=0.9\textwidth]{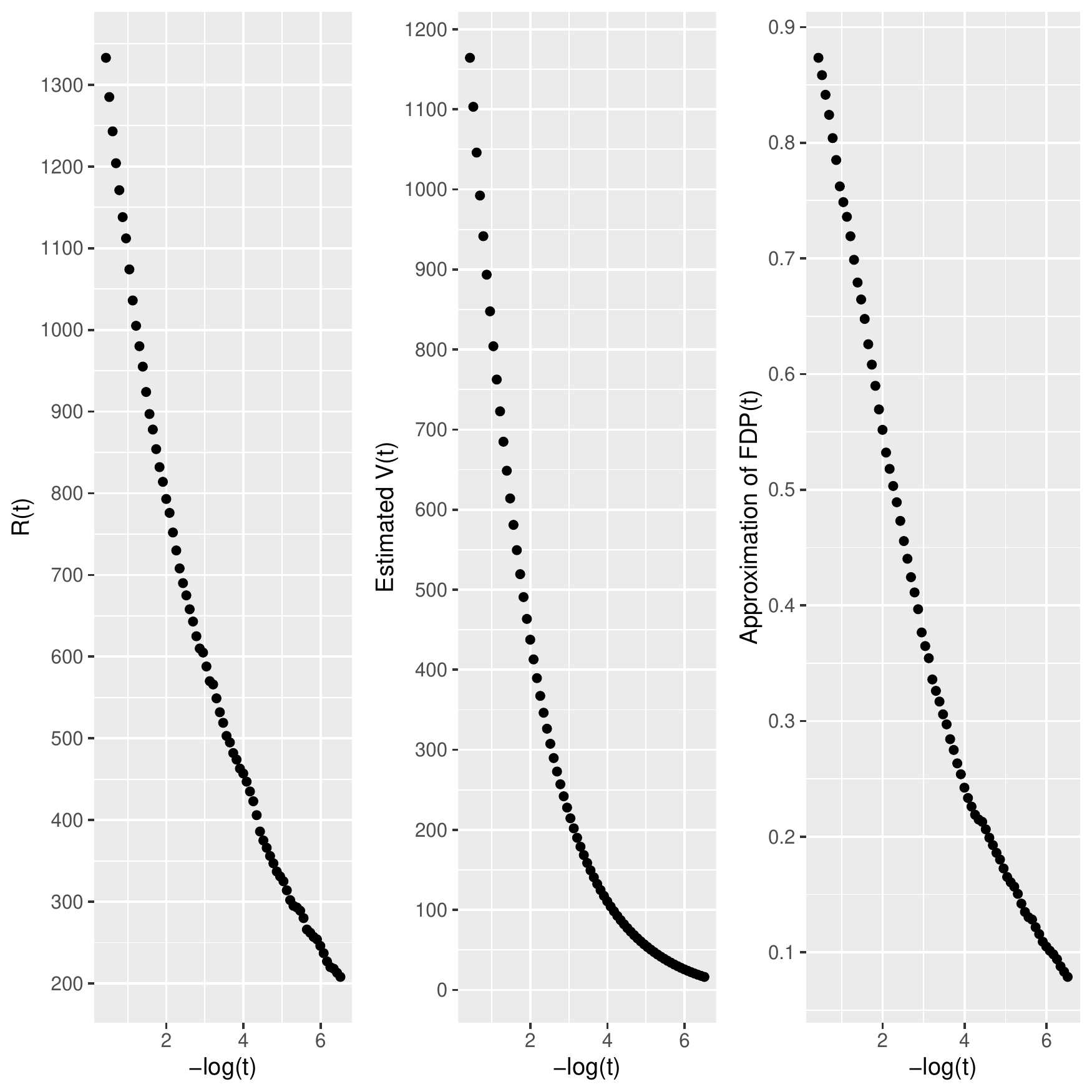}
         \label{fig:sq_panc}
     \end{subfigure}
     \caption{Main results: Overall number of rejections, estimated number of false rejections, and estimated FDP, as functions of the threshold $t$, for both tasks. The values for $t$ on the horizontal axis are plotted on the minus $log_{10}$ scale. }
     \label{fig:main_results}
\end{figure}

The main results of our real data analysis are presented in Figure \ref{fig:main_results}, where we have chosen $k = 3$ for both tasks. According to the number $p = 1{,}579$ and taking into account the magnitude of the observed correlation effects among the $Z$-statistics, a plausible range for the rejection threshold $t$ has been selected for being displayed on the horizontal axes. The subplots in Figure \ref{fig:main_results} illustrate the total number of rejections $(R(t))$, the estimated number of false discoveries $(V(t))$ and the estimated FDP $(\text{FDP}(t))$ over this range. Apparently, all three aforementioned quantities decrease with decreasing $t$ (notice the negative logarithmic scale of the horizontal axes in Figure \ref{fig:main_results}).  For both tasks, $\widehat{\text{FDP}}(t)$ ranges in $[6\%, 17\%]$ for $t \in [10^{-7}, 10^{-5}]$. Table \ref{table:cutoff} tabulates $R(t)$ and $\widehat{\text{FDP}}(t)$ for several threshold values $t$ from the latter range. 

\begin{table}[!htp]
    \caption{\textbf{\bf Main results of the Multi-PFA analysis for both tasks.}}
    \begin{subtable}[c]{0.5\textwidth}
      \centering
        \caption{Pancreatic vs. ADC.}
        	\begin{tabular}{rrr}
		\hline
		Threshold $t$ &$R(t)$ &$\widehat{\text{FDP}}(t)$  \\ 
		\hline
		1.67e-05 & 387 & 0.1619  \\
		7.50e-06 & 356  & 0.1356 \\
		4.12e-06 & 332 & 0.1154\\
		2.76e-06 & 320 & 0.1021\\ 
		5.57e-07 & 275 & 0.0609\\ 
		3.74e-07 & 260 & 0.0540\\ 
		\hline
	\end{tabular}
    \end{subtable}%
     \begin{subtable}[c]{0.5\textwidth}
      \centering
   \caption{Pancreatic vs. SqCC.}
        \begin{tabular}{rrr}
		\hline
		Threshold $t$ &$R(t)$ &$\widehat{\text{FDP}}(t)$  \\ 
		\hline
		1.12e-05 & 331 & 0.1726 \\
		7.50e-06 & 314  & 0.1605 \\
		2.76e-06 & 280 & 0.1303\\
		1.24e-06 & 254 & 0.1092\\ 
		5.57e-07 & 220 & 0.0940\\ 
		3.06e-07 & 208  & 0.0787\\ 
		\hline
	\end{tabular}
    \end{subtable} 
    \label{table:cutoff}
\end{table}
\begin{table}[!htp]
	\caption{\bf Top 10 ranked m/z values based on their original Z-values for the task Pancreatic vs. ADC.} 
        \begin{subtable}{.5\linewidth}
		\raggedright
		\caption{The most sign. m/z-values for Panc.}
		\begin{tabular}{@{\extracolsep{0pt}}lcc}
			\hline\\
			m/z values & \multicolumn{1}{c}{Z-values} & \multicolumn{1}{c}{P-values}  \\
			\hline\\
    886.44 &	-9.917 &  $<10^{-6}$    \\
    840.42 &	-9.570 &  $<10^{-6}$    \\
    679.34 & 	-9.300 &  $<10^{-6}$    \\
    899.45 &	-9.176 &  $<10^{-6}$    \\
    854.42 &	-8.999 &  $<10^{-6}$    \\
    771.38 &	-8.930 &  $<10^{-6}$    \\
    874.43 &	-8.866 &  $<10^{-6}$    \\
    841.42 & 	-8.528 &  $<10^{-6}$    \\
    898.44 &	-8.430 &  $<10^{-6}$    \\
    785.39 &	-8.404 &  $<10^{-6}$    \\
			\hline \\
		\end{tabular}
	\end{subtable}%
		\begin{subtable}{.5\linewidth}
		\caption{The most sign. m/z-values for ADC}
		\raggedright
		\begin{tabular}{@{\extracolsep{0pt}}lcc}
			\hline\\
			m/z values & \multicolumn{1}{c}{Z-values} & \multicolumn{1}{c}{P-values}  \\
			\hline\\
    1647.82 &	8.585 &  $<10^{-6}$    \\
    1575.78	& 8.528 &  $<10^{-6}$    \\
    535.26	&8.300 &  $<10^{-6}$    \\
    521.26	&8.252 &  $<10^{-6}$    \\
    1694.84	&8.245 &  $<10^{-6}$    \\
    1599.79	& 8.197 &  $<10^{-6}$    \\
    531.26	& 8.174 &  $<10^{-6}$    \\
    522.26	 & 7.963 &  $<10^{-6}$    \\
    532.26	&7.844 &  $<10^{-6}$    \\
    529.26&	7.672 &  $<10^{-6}$    \\
			\hline \\
		\end{tabular}
	\end{subtable}%
	\begin{flushleft}
	\end{flushleft}
	\label{table:z10_z1}
\end{table}

\begin{figure}[ht]
  \includegraphics[height = 0.6\textwidth, width=0.9\textwidth]{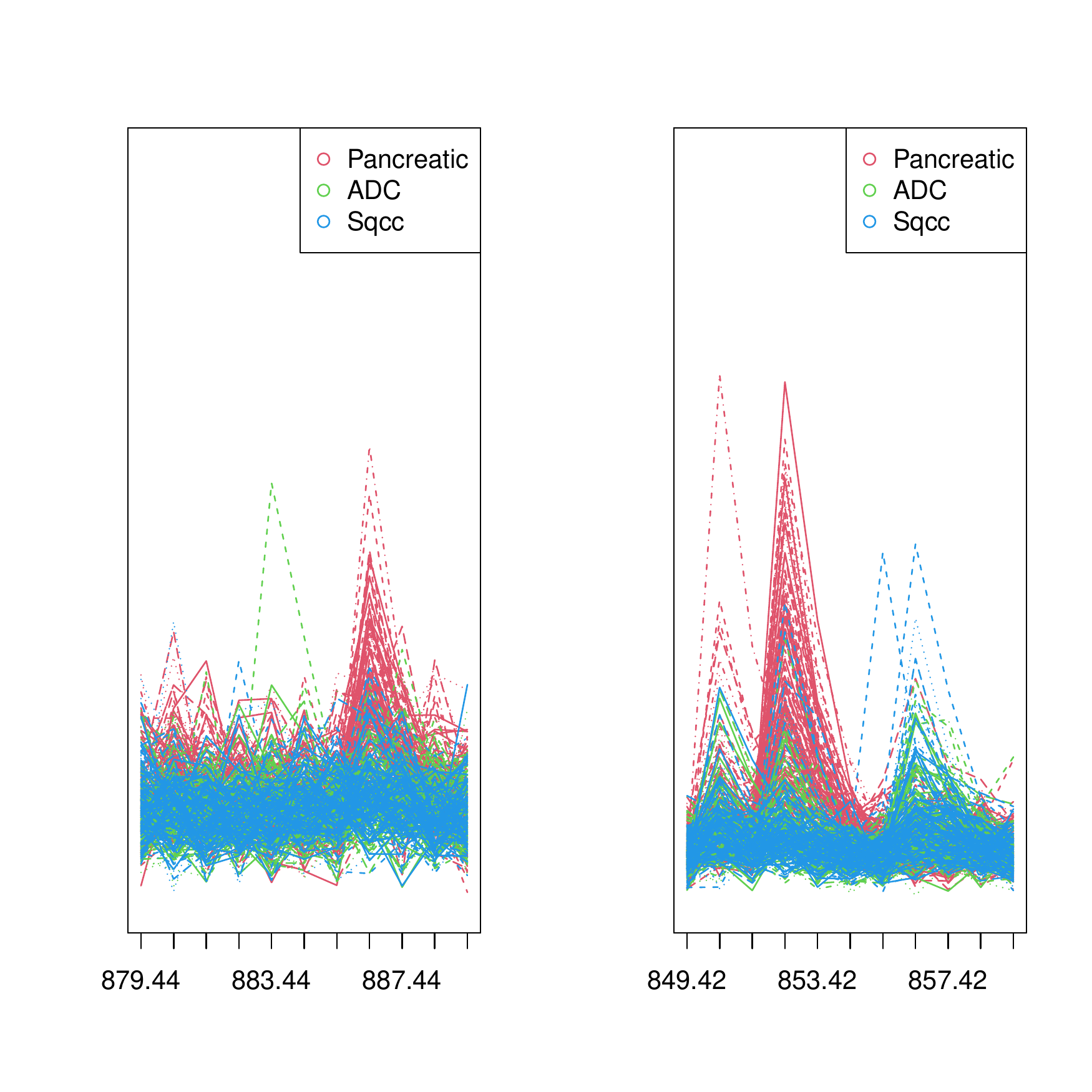}
  \caption{A comparison of the relative intensities at each m/z value across the subtypes for different zooms related to very significant m/z values for the pancreatic association.}
  \label{fig:most_panc}
\end{figure}

\begin{table}[!htp]
	\caption{\bf Top 10 ranked m/z values based on their original Z-values for the task Pancreatic vs. Sqcc.} 
		\begin{subtable}{.5\linewidth}
		\raggedright
		\caption{The most sign. m/z-values for Panc.}
		\begin{tabular}{@{\extracolsep{0pt}}lcc}
			\hline\\
			m/z values & \multicolumn{1}{c}{Z-values} & \multicolumn{1}{c}{P-values}  \\
			\hline\\
771.38 & -9.285 &  $<10^{-6}$    \\
795.39 & -8.839 &  $<10^{-6}$    \\
678.34 & -8.727 &  $<10^{-6}$    \\
785.39 & -8.390 &  $<10^{-6}$    \\
840.42 & -8.334 &  $<10^{-6}$    \\
874.43 & -8.276 &  $<10^{-6}$    \\
886.44 & -8.274 &  $<10^{-6}$    \\
796.39 & -8.266 &  $<10^{-6}$    \\
841.42 & -8.202 &  $<10^{-6}$    \\
854.42 & -8.122 &  $<10^{-6}$    \\
			\hline \\
		\end{tabular}
	\end{subtable}%
		\begin{subtable}{.5\linewidth}
		\caption{The most sign. m/z-values for SqCC}
		\raggedright
		\begin{tabular}{@{\extracolsep{0pt}}lcc}
			\hline\\
			m/z values & \multicolumn{1}{c}{Z-values} & \multicolumn{1}{c}{P-values}  \\
			\hline\\
535.26 &	9.196 &  $<10^{-6}$    \\
1575.78 &	8.860 &  $<10^{-6}$    \\
521.26	&8.664 &  $<10^{-6}$    \\
530.26	&8.406 &  $<10^{-6}$    \\
531.26	&8.381 &  $<10^{-6}$    \\
522.26	&8.339 &  $<10^{-6}$    \\
1694.84	&8.125 &  $<10^{-6}$    \\
1611.80	&7.946 &  $<10^{-6}$    \\
565.28	&7.917 &  $<10^{-6}$    \\
1764.87	&7.910 &  $<10^{-6}$    \\
			\hline \\
		\end{tabular}
\end{subtable}%
	\begin{flushleft}
	\end{flushleft}
	\label{table:z10_z2}
\end{table}
 To the best of our knowledge, no biomarkers have been confirmed yet for this dataset (pancreas vs. lung task).\cite{behrmann} However, according to the previous reports\cite{behrmann, boskamp}, m/z values at 836.5 Da, 852.4 Da and 868.5 Da might be potential biomarkers. Since we have used data based on a different resolution (0.4 Da) compared to the aforementioned reports, we identified as statistically significant m/z-values which are closely related to the previously published insights. Namely, the adjacent molecules have appeared to be highly statistically significant. In particular, for the m/z-value 836.5 Da, based on the spectral filtering used in our analysis, adjacent m/z values (836.42, 837.41 and 838.42) have been indicated as highly significant for both tasks. Similarly, for the m/z value 852.4 Da,  adjacent m/z values have occurred as highly considerable (853.42 and 854.42). Similarly, we observed for the m/z value at 868.50 the adjacent m/z value 869.43. Figure \ref{fig:most_panc} illustrates the identified peaks related to the pancreatic status. The Multi-PFA method indicates m/z values around 886.42 Da and 852.42 Da as highly statistically significant. Tables \ref{table:z10_z1} and \ref{table:z10_z2} tabulate the 10 top-ranked m/z values for both tasks. Their corresponding $Z$-values lie in the very tails of the empirical distributions plotted in Figure \ref{fig:emp_z_values}. Negative signs in Tables \ref{table:z10_z1} and \ref{table:z10_z2} indicate those m/z values that are distinctive for the pancreatic cancer subtype. 

\section{Discussion and Outlook}\label{sec-discussion}
Motivated by MALDI association studies, this study's objective has been to evaluate the strength of associations between a nominal variable of interest \textendash \, describing cancer subtypes in our application \textendash \, and a large number of measured variables (in our case given by m/z values). We have proposed an approach to screen all features and identify the most associative ones with the multi-class outcome. Our approach decomposes the outcome variable into multiple baseline-category pairs, and approximates the false discovery proportion under arbitrary correlation dependency within each pair. As demonstrated on simulated data, this approach leads to a sensible balance between the (average) numbers of true and false rejections. 
Furthermore, both our simulation study and the presented application to real data demonstrate that the proposed method can be applied even in cases where the sample size is considerably smaller than the number of features. This makes the procedure attractive for medical screening applications. 
Two central assumptions underlie the proposed methodology: First, we assume sparsity in the sense that the number of active features is small. Second, we assume that the dependency structure amongst the considered $Z$-statistics can accurately be approximated by a multi-factor model. The approach of MMM, however, is flexible in the sense that it does not depend on heavy assumptions. For its applicability, we only have to assume that the multi-class outcome is associated with features (m/z values), and that this association can accurately be described by a (marginal) multinomial regression model for each m/z-value separately. 

From the application perspective, we have applied the Multi-PFA method to a MALDI imaging dataset consisting of a large number of m/z values and one spectrum from each patient. To the best of our knowledge, we are the first to address the three-class problem for this dataset. In this way, we have provided a more detailed analysis for both lung cancer subtypes than previous studies. 

There are several potential directions for further research. First, it might be captivating to consider different supervised learning methods (for instance, a neural network with more than one layer) instead of the multinomial regression model proposed in this paper. Second, it may be of interest  to evaluate the uncertainty regarding the realized FDP in order to provide a confidence region for it, in addition to mere point estimation of the FDP. Third, disentangling correlation effects and regression effects on the empirical $Z$-statistics distribution in a more detailed manner is an interesting follow-up research topic in our context.


\section*{Acknowledgments}
The authors sincerely thank Johannes Leuschner for his invaluable aid in performing the data-processing steps of MALDI data in Matlab R2018. 

\subsection*{Author contributions}
VV has performed data modelling and analysis, as well as R programming. TD has conceptualized the research project and proposed the usage of MMM. Both authors have written the manuscript together.

\subsection*{Financial disclosure}
The authors gratefully acknowledge financial support from the German Research Foundation within the framework of RTG 2224, entitled "$\pi^{3}$: Parameter Identification—Analysis, Algorithms, Applications".

\subsection*{Conflict of interest}
The authors declare no potential conflict of interests.










\end{document}